\documentclass[aps,twocolumn,showkeys,reprint,amsmath,amssymb,prb]{revtex4-1}

\usepackage{graphicx,color}
\usepackage{dcolumn}
\usepackage{hyperref}
\usepackage{bm}
\usepackage{stmaryrd}
\usepackage{latexsym}
\usepackage{amssymb}
\usepackage{amsfonts}
\usepackage{amsmath}
\usepackage{epstopdf}

\begin{document}

\title{Coexistence of ferromagnetism and superconductivity at KTaO$_3$ heterointerfaces}

\author{Zhongfeng Ning,$^{1}$ Jiahui Qian,$^{1}$ Yixin Liu,$^{2,3}$ Fan Chen,$^{2,3}$ Mingzhu Zhang,$^{2,3}$ Liwei Deng,$^{2,3}$ Xinli Yuan,$^{4}$ Qingqin Ge,$^{4}$ Hua Jin,$^{2}$ Guanqun Zhang,$^{1}$ Wei Peng,$^{2,3}$ Shan Qiao,$^{2,3,*}$ Gang Mu,$^{2,3,*}$ Yan Chen$^{1,*}$ and Wei Li$^{1,*}$}

\affiliation
{$^1$State Key Laboratory of Surface Physics and Department of Physics, Fudan University, Shanghai 200433, China\\
 $^2$National Key Laboratory of Materials for Integrated Circuits, Shanghai Institute of Microsystem and Information Technology, Chinese Academy of Sciences, Shanghai 200050, China\\
 $^3$University of Chinese Academy of Sciences, Beijing 100049, China\\
 $^4$Thermo Fisher Scientific China, Shanghai 201203, China
 }

\date{\today}

\begin{abstract}
\noindent\textbf{ABSTRACT:} The coexistence of superconductivity and ferromagnetism is a long-standing issue in superconductivity due to the antagonistic nature of these two ordered states. Experimentally identifying and characterizing novel heterointerface superconductors that coexist with magnetism presents significant challenges. Here, we report the experimental observation of two-dimensional long-range ferromagnetic order in the KTaO$_3$ heterointerface superconductor, showing the coexistence of superconductivity and ferromagnetism. Remarkably, our direct current superconducting quantum interference device measurements reveal an in-plane magnetization hysteresis loop persisting above room temperature. Moreover, the first-principles calculations and X-ray magnetic circular dichroism measurements provide decisive insights into the origin of the observed robust ferromagnetism, attributing it to oxygen vacancies that localize electrons in nearby Ta 5$d$ states. Our findings not only suggest KTaO$_3$ heterointerfaces as time-reversal symmetry breaking superconductors, but also inject fresh momentum into the exploration of the intricate interplay between superconductivity and magnetism, enhanced by the strong spin-orbit coupling inherent to the heavy Ta in 5$d$ orbitals of KTaO$_3$ heterointerfaces.
\end{abstract}


\maketitle

\noindent\textbf{INTRODUCTION}

\noindent Two-dimensional oxide heterointerfaces exhibit a wide range of emergent quantum phenomena inaccessible in their bulk individuals due to the strong interplay between electrons with Coulomb interaction and the interfacial electron-phonon coupling at the interfaces~\cite{ref1,ref2,ref3}. Among them, a landmark oxide interface is LaAlO$_3$/SrTiO$_3$~\cite{ref4}, which displays a plethora of appealing physical properties, encompassing two-dimensional electron gases with high electron mobility~\cite{ref5}, strong Rashba-like spin-orbit coupling~\cite{ref7,ref8,Soumyanarayanan}, interfacial superconductivity~\cite{ref6,Caviglia}, and ferromagnetism~\cite{ref9,ref10}. In particular, the coexistence of superconductivity and ferromagnetism has also been unveiled by high-resolution magnetic torque magnetometry~\cite{ref11}, scanning superconducting quantum interference device (SQUID)~\cite{ref12}, and magnetoresistance~\cite{ref13}. This suggests the presence of a nontrivial superconducting phase in the ground state, such as an intriguing Fulde-Ferrell-Larkin-Ovchinikov-type condensate of Cooper pairs with finite momentum~\cite{ref18} and a mixed-parity superconducting phase with an admixture of spin-singlet and spin-triplet pairing components~\cite{LFu2015,Nagaosa2012} that is a candidate platform for realizing Majorana modes~\cite{Potter2011}. However, the experimental challenges posed by the extremely low superconducting critical temperature $T_{c}$ (below 250 mK) of SrTiO$_3$ heterointerfaces have hindered a comprehensive exploration of the underlying properties~\cite{ref6}, leaving the origin of these quantum phases still elusive~\cite{Fitzsimmons,ref22} and motivating the search for a more favorable heterointerface superconductor that hosts the intricate interplay between ferromagnetism and superconductivity.

\begin{figure*}
\centering
\includegraphics[bb=15 495 585 780,width=\textwidth]{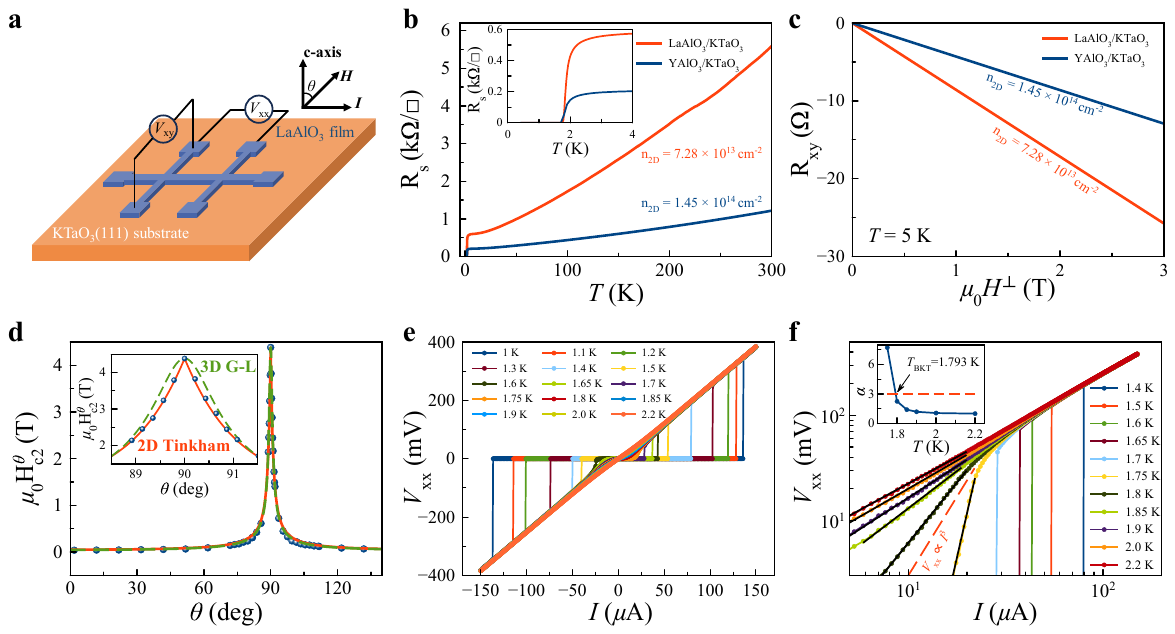}
\caption{Two-dimensional superconductivity of KTaO$_3$ heterointerfaces. (a) Schematic structure of the Hall bar device on a representative LaAlO$_3$/KTaO$_3$(111). (b) Longitudinal electrical resistance $\mathrm{R}_{\mathrm{s}}$ as a function of temperature for LaAlO$_3$ and YAlO$_3$ films on KTaO$_3$(111) substrates. Inset: Magnified view of the low temperature regime. (c) Corresponding perpendicular field $\mu_0H^{\perp}$-dependent transverse Hall resistance $\mathrm{R}_{\mathrm{xy}}$ measured at the temperature of 5 K. (d) Out-of-plane polar angular $\theta$ dependence of $\mu_0\mathrm{H}_{\mathrm{c2}}^{\theta}$ for LaAlO$_3$/KTaO$_3$(111). Inset: Magnified view around $\theta=90^{\circ}$. The red solid and green dotted lines represent theoretical fits using the two-dimensional (2D) Tinkham and the three-dimensional (3D) anisotropic Ginzburg-Landau (G-L) models, respectively. (e) Temperature-dependent $I$-$V_{\mathrm{xx}}$ measurements for LaAlO$_3$/KTaO$_3$(111). (f) Corresponding logarithmic scale representation of (e). The red dashed line indicates $V_{\mathrm{xx}} \varpropto I^3$. Inset: The extracted power-law fitting exponent $\alpha$ as a function of $T$. The $T_{\mathrm{BKT}}=$ 1.793 K is defined by $\alpha = $ 3.}
\label{fig1}
\end{figure*}

Very recently, an unforeseen crystalline-orientation-dependent superconductivity has been observed at KTaO$_3$ heterointerfaces with $T_\text{c}$ as high as 2 K~\cite{ref27,ref23,ref25,ref26}. This is nearly one order of magnitude higher in $T_\text{c}$ than that of SrTiO$_3$~\cite{ref6}. The experimental findings also demonstrate that electronic states near the Fermi energy derived from Ta 5$d$ orbitals with strong spin-orbit coupling play a decisive role in electronic conduction at the KTaO$_3$ heterointerfaces~\cite{Bruno2019,Mallik,Hua2022,Arnault,Ahadi2023}. Moreover, the existence of intrinsic anomalous Hall effect observed by electrical transport measurements at non-superconducting KTaO$_3$ heterointerfaces~\cite{Tawhid2022,Krantz2022} suggests an emergent ferromagnetism with time-reversal symmetry breaking. Additionally, at superconducting KTaO$_3$ heterointerfaces, the in-plane azimuthal angle-dependent magnetoresistance and the superconducting critical field exhibit striking twofold symmetric oscillations deep inside the superconducting phase, whereas the anisotropy vanishes in the normal phase, as highlighted in our recent notable study~\cite{ref26}. This observation indicates the intrinsic nature of the mixed-parity superconducting ground state with an admixture of $s$-wave and $p$-wave pairing components inherent to the inversion symmetry breaking at the KTaO$_3$ heterointerfaces~\cite{ref26}. Although the component of $p$-wave pairing could be theoretically stabilized by the long-range ferromagnetic order~\cite{ref32,Zou2020}, the existence of ferromagnetism at superconducting KTaO$_3$ heterointerfaces, akin to the observation at superconducting SrTiO$_3$ heterointerfaces~\cite{ref11,ref12,ref13}, is highly desired to examine by using direct experimental accesses.

In this work, utilizing the measurements of direct current SQUID, we observe a distinct signal of an in-plane ferromagnetic hysteresis loop in the KTaO$_3$ heterointerface superconductor, attributing it to intrinsic ferromagnetism that persists above room temperature. Moreover, the first-principles calculations and X-ray magnetic circular dichroism (XMCD) measurements provide decisive evidences in support of the robust ferromagnetism of superconducting KTaO$_3$ heterointerfaces, which derives from the local moments of the Ta$^{4+}$: $5d^1$ ions brought about by the oxygen vacancies that host the two-dimensional electron gases formed at the KTaO$_3$ heterointerfaces, which are responsible for the emergence of interfacial superconductivity. Remarkably, these findings indicate the coexistence of ferromagnetism and superconductivity at the KTaO$_3$ heterointerfaces, offering novel insights into the intricate superconducting properties at these interfaces.

\notag\

\noindent\textbf{RESULTS AND DISCUSSION}

\noindent Both the LaAlO$_3$ and YAlO$_3$ thin films are grown by pulsed laser deposition on top of the single-crystalline KTaO$_3$(111) substrates (see Methods in Supporting Information). Atomic force microscopy (AFM) characterization reveals that the surface of KTaO$_3$ substrates and thin films are atomically flat (see Figure S1, YAlO$_3$ shown in Ref.~\onlinecite{ref26}), suggesting a high-quality growth of the thin films on KTaO$_3$. On the other hand, X-ray diffraction (XRD) exposes the absence of epitaxial Bragg reflection peaks in both the LaAlO$_3$ and YAlO$_3$ thin films (see Figure S2, YAlO$_3$ presented in Ref.~\onlinecite{ref26}), confirming the deposition of an amorphous phase for both the LaAlO$_3$ and YAlO$_3$ thin films on the KTaO$_3$(111) substrates. These results align well with previous investigations~\cite{ref23,ref25,ref26}. Subsequently, we conduct electrical transport measurements on the devices of the as-grown thin films arranged in Hall bar configurations (see Figure~\ref{fig1}a)~\cite{ref26,XueH2022}.

\begin{figure*}
\centering
\includegraphics[bb=12 375 570 620,width=\textwidth]{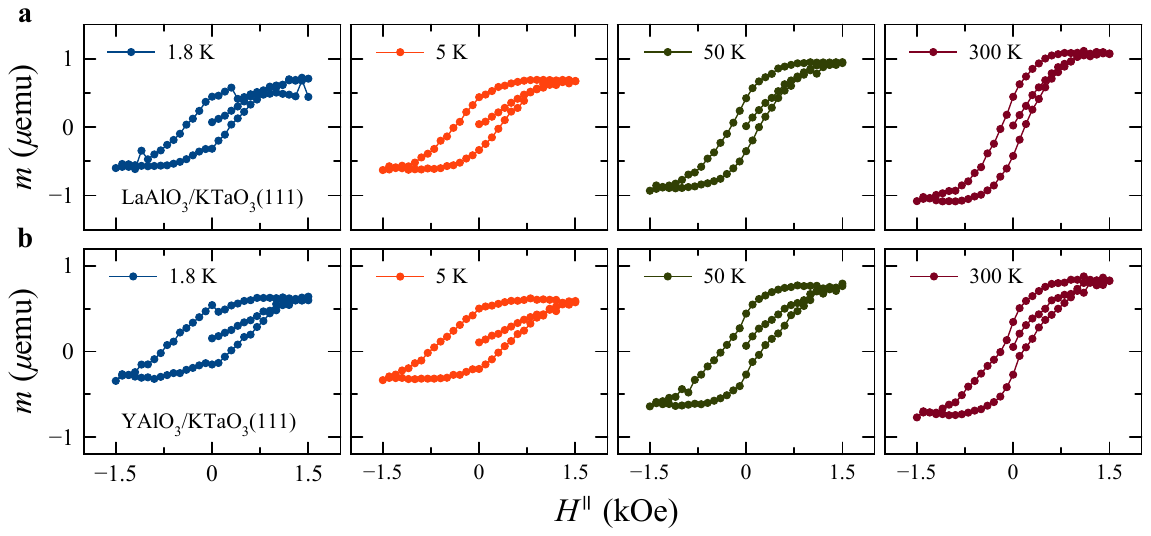}
\caption{Ferromagnetic properties of KTaO$_3$ heterointerfaces. Temperature-dependent in-plane ferromagnetic hysteresis loops ($m$-$H^{\parallel}$) for (a) LaAlO$_3$/KTaO$_3$(111) and (b) YAlO$_3$/KTaO$_3$(111), corrected for the diamagnetic contribution of the bare KTaO$_3$ substrate. The applied magnetic field $H^{\parallel}$ ranges from -1.5 kOe to +1.5 kOe, with the field parallel to the surface of thin films.}
\label{fig2}
\end{figure*}

The longitudinal electrical resistances $\mathrm{R}_{\mathrm{s}}$ in Figure~\ref{fig1}b exhibit metallic behavior from room temperature to 2.5 K for both LaAlO$_3$/KTaO$_3$(111) and YAlO$_3$/KTaO$_3$(111), indicating the formation of two-dimensional electron gases at their heterointerfaces, given that their bulk constituents are nonmagnetic insulators. In addition, the charge carriers in both cases are electrons with an estimated charge carrier density of $7.28\times 10^{13}$ cm$^{-2}$ and $1.45\times 10^{14}$ cm$^{-2}$, respectively, by the Hall measurements at 5 K (see Figure~\ref{fig1}c). Remarkably, the superconducting transitions start at $T_{\mathrm{c}}^{\mathrm{onset}}=$ 2.32 K and 2.39 K, and resistances drop to zero at $T_{\mathrm{c}}^{\mathrm{zero}}=$ 1.66 K and 1.7 K, respectively, for LaAlO$_3$/KTaO$_3$ and YAlO$_3$/KTaO$_3$. In the LaAlO$_3$/KTaO$_3$(111), the field-dependent $\mathrm{R}_{\mathrm{s}}$ shown in Figure S3 yields the upper critical fields $\mu_0\mathrm{H}_{\mathrm{c2}}^{\parallel}(0.6~\mathrm{K}) =$ 5 T for fields parallel to the sample plane surface and $\mu_0\mathrm{H}_{\mathrm{c2}}^{\perp}(0.6~\mathrm{K}) =$ 0.45 T for fields parallel to the crystallographic $c$-axis. Here, $\mu_0\mathrm{H}_{\mathrm{c2}}$ is defined as the magnetic field at the midpoint of the electrical resistance transition. The extracted temperature dependence of $\mu_0\mathrm{H}_{\mathrm{c2}}$ is shown in Figure S3c, where a large ratio $\mathrm{H}_{\mathrm{c2}}^{\parallel}/\mathrm{H}_{\mathrm{c2}}^{\perp}$ is perceived, suggesting strong anisotropic superconductivity of LaAlO$_3$/KTaO$_3$(111). Quantitatively, the out-of-plane polar angular $\theta$-dependent $\mu_0\mathrm{H}^{\theta}_{\mathrm{c2}}$ at 1.6 K is used to verify this intriguing behavior, as shown in Figure~\ref{fig1}d. Using the two-dimensional Tinkham and the three-dimensional anisotropic Ginzburg-Landau models to fit the $\mu_0\mathrm{H}^{\theta}_{\mathrm{c2}}$, given by $\frac{\mathrm{H}_{\mathrm{c2}}^{\theta}|\cos\theta|}{\mathrm{H}_{\mathrm{c2}}^{\perp}} + (\frac{\mathrm{H}_{\mathrm{c2}}^{\theta}\sin\theta}{\mathrm{H}_{\mathrm{c2}}^{\parallel}})^2 = 1$ and $(\frac{\mathrm{H}_{\mathrm{c2}}^{\theta}\cos\theta}{\mathrm{H}_{\mathrm{c2}}^{\perp}})^2 + (\frac{\mathrm{H}_{\mathrm{c2}}^{\theta}\sin\theta}{\mathrm{H}_{\mathrm{c2}}^{\parallel}})^2 = 1$, respectively~\cite{Tinkham1963,ref38}, a cusp-like peak is clearly observed at around $\theta=90^{\circ}$ (see Figure~\ref{fig1}d, inset), which is well described by the two-dimensional Tinkham model, as frequently observed in heterointerface superconductivity~\cite{ref38,ref26,Zhang2022} and layered transition metal dichalcogenides~\cite{LuJM,ref32}. Qualitatively similar results have also been observed in YAlO$_3$/KTaO$_3$~\cite{ref26}, unambiguously demonstrating the intrinsic two-dimensional superconductivity at the KTaO$_3$ heterointerfaces.

To further examine the intrinsic interfacial superconductivity at the KTaO$_3$ heterointerfaces, we conduct measurements of the current-voltage ($I$-$V_{\mathrm{xx}}$) characteristics at various temperatures near $T_c$ (see Figure~\ref{fig1}e). These characteristics, depicted on a log-log scale in Figure~\ref{fig1}f, reveal a distinct critical current $I_c$ of 79.1 $\mu$A at 1.4 K, below $T_c$. The $I_c$ gradually decreases with increasing $T$, eventually vanishing, indicating a transition from superconducting to normal phases. In the normal phase, we have $V_{\mathrm{xx}} \varpropto I$ according to Ohm's law. However, a steeper power law $V_{\mathrm{xx}} \varpropto I^{\alpha(T)}$ emerges as $T$ drops. Following the Berezinskii-Kosterlitz-Thouless (BKT) definition~\cite{Kosterlitz,Beasley}, the transition temperature $T_{\mathrm{BKT}}$ corresponds to the dissociation of vortex-antivortex pairs. This obeys the universal scaling relation $V_{\mathrm{xx}} \varpropto I^{3}$, since the superconducting phase consists of bound vortex-antivortex pairs in two-dimensional superconductors. We determine $T_{\mathrm{BKT}}=$ 1.793 K from where $\alpha=$ 3 interpolates. Furthermore, $T_{\mathrm{BKT}}$ can be alternatively evaluated from the formula $\mathrm{R}_\text{s}=R_\text{0}$exp$[-b(T/T_\text{BKT}-1)^{-1/2}]$, where $R_\text{0}$ and $b$ are material parameters~\cite{Halperin}. Applying such a theoretical fit to the measured $\mathrm{R}_{\mathrm{s}}$ yields $T_{\mathrm{BKT}}=$ 1.878 K (see Figure S4). As expected, the $T_{\mathrm{BKT}}$ obtained from these two independent approaches is close to $T_{\mathrm{c}}^{\mathrm{zero}}$, providing additional evidence for the two-dimensional superconducting nature of the KTaO$_3$ heterointerfaces.

\begin{figure*}[tp]
\centering
\includegraphics[bb=5 535 510 820,width=\textwidth]{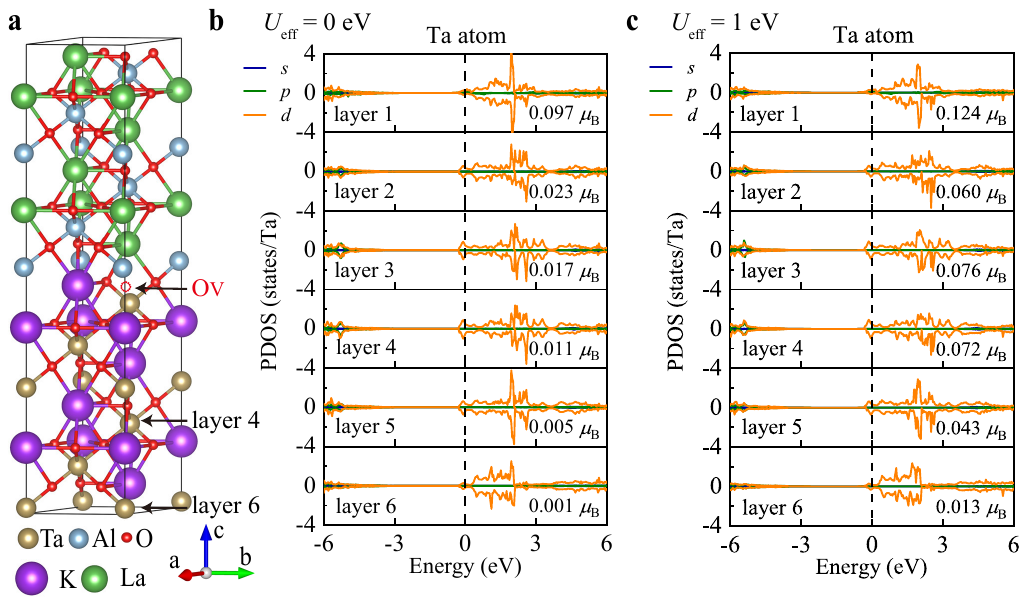}
\caption{First-principles calculations on the KTaO$_3$ heterointerface. (a) Schematic structure of the superlattices (LaAlO$_3$)$_6$/(KTaO$_3$)$_6$(111) with the presence of an oxygen vacancy. The red dotted circle denotes the oxygen vacancy (O$_{\mathrm{V}}$) at the heterointerface of LaAlO$_3$/KTaO$_3$(111). The PDOS for Ta atoms, calculated by using the GGA+$U_{\mathrm{eff}}$ method with (b) $U_{\rm eff}$ = 0 eV and (c) $U_{\rm eff}$ = 1.0 eV. The numbers listed in the figure represent the magnetization of Ta in each layer. Here, we set the Fermi energy to zero.}
\label{fig3}
\end{figure*}

Next, we delve into the magnetic properties of KTaO$_3$ heterointerfaces. Theoretically, the two-dimensional conducting electron gases at the interfaces of KTaO$_3$ heterostructures primarily originate from the partially filled 5$d$ orbitals of the Ta atoms induced by oxygen vacancies~\cite{ref25,ref26}, these vacancies themselves have the potential to induce ferromagnetism. Experimentally, the zero-field-cooling (ZFC) and field-cooling (FC) curves of the as-grown thin films of LaAlO$_3$ and YAlO$_3$, measured at a temperature between 1.8 and 300 K with an applied in-plane field ($H^{\parallel}$) of 50 Oe, are summarized in Figure S5. The figures exhibit the pronounced separation between the ZFC and FC curves up to 300 K, showing a signature of in-plane ferromagnetism at the superconducting KTaO$_3$ heterointerfaces that persists at room temperature. In Figure~\ref{fig2}, the well-defined magnetization-magnetic field ($m\sim H^{\parallel}$) hysteresis loops observed at different temperatures for the LaAlO$_3$/KTaO$_3$ and YAlO$_3$/KTaO$_3$ provide clear evidence of the temperature dependence of long-range ferromagnetic order. Control experiments on an as-received KTaO$_3$ substrate without LaAlO$_3$ or YAlO$_3$ overlayers only show a diamagnetic background (see Figures S6 and S7). Additionally, we conduct X-ray photoelectron spectroscopy (XPS) and scanning electron microscopy (SEM)-energy dispersive X-ray spectroscopy (EDS) measurements to actively eliminate concerns related to extrinsic magnetic contamination during sample preparation. As anticipated, the samples exhibit no evidence of magnetic elements, such as Fe, Co, Mn, Ni or Cr, as shown in Figures S8 and S9. As a consequence, we attribute the universally and consistently observed distinct ferromagnetic signal depicted in Figure~\ref{fig2} to the intrinsic nature of the KTaO$_3$ heterointerfaces. Specifically, we attribute the emergence of two-dimensional ferromagnetism to the Ta$^{4+}: 5d^1$ electrons induced by the high concentration of oxygen vacancies at the KTaO$_3$ heterointerfaces~\cite{ref25}. This observation is noteworthy because their bulk counterparts are typical nonmagnetic insulators. It is also important to note that the distinction contrasts with a previous report, in which the ferromagnetic response originates from an extrinsic magnetic thin film of EuO~\cite{ref42,NPChen2024}.

\begin{figure*}%[tbhp]
\centering
\includegraphics[bb=10 400 580 610,width=\textwidth]{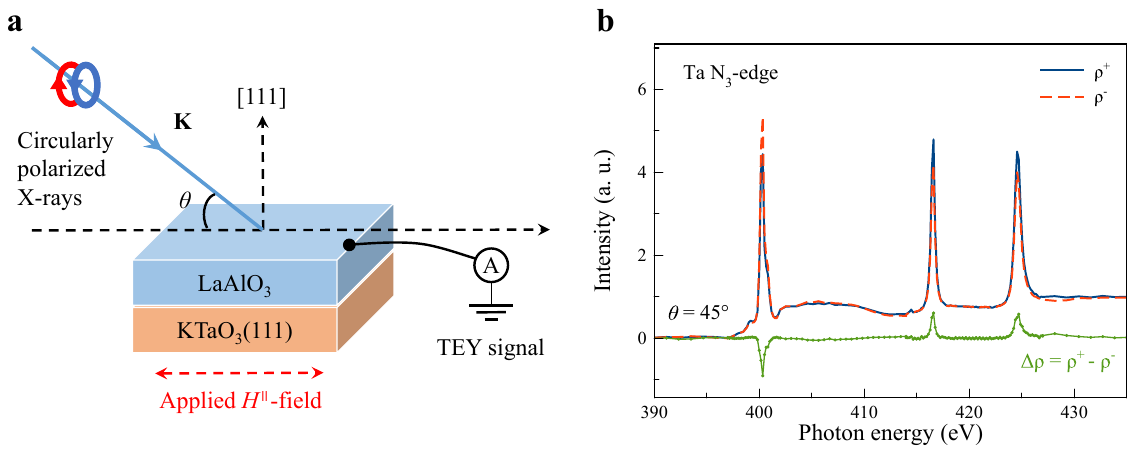}
\caption{XMCD measurement on a LaAlO$_3$/KTaO$_3$(111) heterostructure. (a) Schematic illustration of the experimental configuration for the XMCD measurement, where $\mathbf{K}$ represents the x-ray propagation vector. (b) Spectrum of the XMCD at Ta N$_3$ absorption edge observed in the in-plane geometry with a fixed incidence angle $\theta= 45^{\circ}$, revealing ferromagnetic Ta at $T$ = 15 K.}
\label{fig4}
\end{figure*}

Furthermore, we actively compare the in-plane and out-of-plane $m\sim H$ loop SQUID measurements of the LaAlO$_3$/KTaO$_3$ and YAlO$_3$/KTaO$_3$ (see Figures S10-S13). These measurements indicate that the temperature dependence of the long-range ferromagnetic order aligns with the in-plane orientation of the KTaO$_3$ heterointerfaces, a consistency observed with the results from the first-principles calculations (see also the first-principles calculations below for a comprehensive analysis) and the anomalous Hall effect and the in-plane field-dependent magnetoresistance with butterfly-like hysteresis (see Figure S14). In addition, we quantify the in-plane magnetization of the itinerant electron gases formed at the KTaO$_3$ heterointerfaces, revealing a magnitude of approximately $\simeq$0.2 $\mu_{\mathrm{B}}$ per interface unit cell (if we attribute all the magnetization to the interface). It is interesting to indicate that the KTaO$_3$ heterointerfaces display a non-monotonic behavior in temperature-dependent magnetization (see Figure~\ref{fig2} and Figure S15). This non-monotonic pattern, also observed in previously explored LaAlO$_3$/SrTiO$_3$~\cite{ref10} and VSe$_2$ monolayers on van der Waals substrates~\cite{Bonilla}, is attributed to the structural phase and charge-density-wave transitions, respectively. Since KTaO$_3$ shares a number of properties in common with SrTiO$_3$~\cite{ref26}, the observed non-monotonic temperature-dependent magnetization at KTaO$_3$ heterointerfaces may be also attributed to the structural phase transitions located at the heterointerfaces. This non-monotonicity underscores the critical role of the KTaO$_3$ surface and its phase transition in the origin of these emergent magnetic properties, and LaAlO$_3$ or YAlO$_3$ overlayers significantly amplify these effects by forming two-dimensional electron gases at the KTaO$_3$ heterointerfaces~\cite{ref10}. Further experimental efforts would be needed to confirm this underlying nature of the observed non-monotonicity in temperature-dependent magnetization. Moreover, at the lowest temperature of this measurement limited $T$ = 1.8 K, which is also lower than $T_{\mathrm{c}}^{\mathrm{onset}}$ [2.32 K for LaAlO$_3$/KTaO$_3$(111) and 2.39 K for YAlO$_3$/KTaO$_3$(111)], the conspicuous signal of in-plane ferromagnetic hysteresis loops is reproducibly observed in the superconducting transition regime (see Figure~\ref{fig2}), and the magnetization does not seem to be expelled entirely below $T_c$. These universal and striking results are strongly reminiscent of the superconducting KTaO$_3$ heterointerfaces spatially coexisting ferromagnetic order~\cite{Pfleiderer} (see also Figure S14). Therefore, we experimentally suggest the coexistence of ferromagnetism and superconductivity at the KTaO$_3$ heterointerfaces, which has also long been a topic of interest sought in nanoscience and nanotechnology communities~\cite{Ghosh}.

To deepen our understanding of the intrinsic ferromagnetism emerging at the KTaO$_3$ heterointerfaces, we delve into the electronic and magnetic properties using the first-principles calculations (see Methods in Supporting Information). For simplicity, we establish a theoretical model with the superlattice (KTaO$_3$)$_6$/(LaAlO$_3$)$_6$(111) comprising a supercell of six layers each for LaAlO$_3$ and KTaO$_3$, despite the amorphous phase of LaAlO$_3$ thin films grown on KTaO$_3$(111) substrates (see Figure S2), as schematically illustrated in Figure S16a. Interestingly, in the stoichiometric heterostructure (KTaO$_3$)$_6$/(LaAlO$_3$)$_6$(111), devoid of oxygen vacancy, the calculated layer-dependent projected density-of-states (PDOS) reveal that the system is a nonmagnetic band insulator with an energy band gap of $\simeq$3.0 eV (see Figure S16), consistent with previous studies~\cite{Fujii1976}. This result indicates that the emergent two-dimensional electron gases at the KTaO$_3$ heterointerfaces are unlikely to stem from the diverging Coulomb field due to the polarity of individual layers of LaAlO$_3$ and KTaO$_3$, but rather from the presence of oxygen vacancies.

\begin{figure*}%[tbhp]
\centering
\includegraphics[bb=5 10 785 560,width=\textwidth]{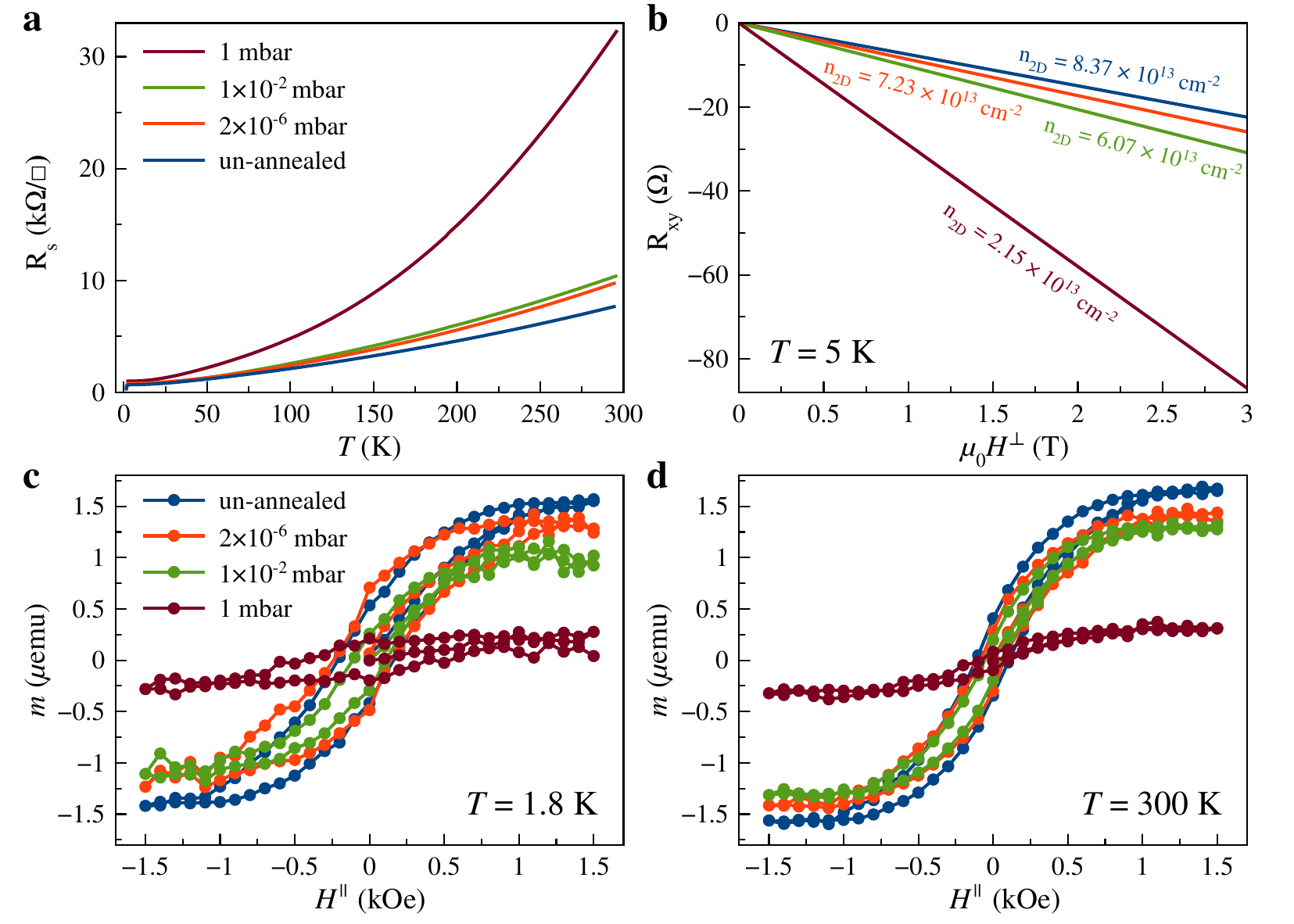}
\caption{Influence of post-annealing processes. (a) R$_{\mathrm{s}}$ as a function of $T$ for LaAlO$_3$/KTaO$_3$(111) subjected to post-annealing processes in various oxygen partial pressures of 1, $1\times10^{-2}$, and $2\times10^{-6}$ mbar at 400 $^{\circ}$C for 1 h. We also include the data from the un-annealed LaAlO$_3$/KTaO$_3$(111) sample. (b) Corresponding transverse Hall resistance R$_{\mathrm{xy}}$ as a function of $\mu_0H^{\perp}$ at 5 K. In-plane magnetization measured at (c) 1.8 K and (d) 300 K for films subjected to post-annealing processes. It is important to note that the diamagnetic background from the KTaO$_3$ substrate has been subtracted on the high field slope at each measured temperature.}
\label{fig5}
\end{figure*}

Principally, each oxygen vacancy donates two electrons into the LaAlO$_3$/KTaO$_3$(111) interface. In turn, the induced interface charges are typically spread over adjacent layers, giving rise to a Ta$^{4+}$$\sim$O$^{2-}$$\sim$Ta$^{5+}$-like electronic configuration, which corresponds to several partially filled Ta 5$d$ sub-bands that can lead to interfacial ferromagnetism. As expected, density functional theory calculations show that in the presence of oxygen vacancies (O$_{\mathrm{V}}$, denoted by red dotted circle shown in Figure~\ref{fig3}a), the system behaves as a ferromagnetic metal with the conducting electrons mainly derived from the Ta 5$d$ orbitals partially hybridized with the oxygen 2$p$ orbitals (see Figure S17), which are responsible for the emergence of two-dimensional superconductivity~\cite{ref23,Bruno2019}. Furthermore, Figure~\ref{fig3}b,c show the spin-resolved PDOS for Ta atoms associated with the value of the magnetic moments of each Ta layer. The calculated magnetic moment of Ta, brought about by the nearby oxygen vacancy, is about 0.124 $\mu_{\mathrm{B}}$ (see Figure~\ref{fig3}c). These calculated results align qualitatively and quantitatively with the magnetization measurements presented in Figure~\ref{fig2}, affirming the intrinsic nature of the long-range ferromagnetic order at the KTaO$_3$ heterointerface. Additionally, it is noteworthy that we have adjusted the electron correlation effect of Ta 5$d$ orbitals through the simplified on-site repulsion strength $U_{\mathrm{eff}}$ on Ta 5$d$ orbitals~\cite{Pentcheva}. This adjustment yields a magnetic moment that closely corresponds to the experimental findings (see Figure~\ref{fig3}c and Figure S18).

Experimentally, we independently conduct synchrotron XMCD measurements (see Methods in Supporting Information) to further investigate whether the interfacial ferromagnetism stems from the intrinsic spin polarization of Ta 5$d$ electrons at the KTaO$_3$ heterointerfaces, utilizing the element-specific technique and high sensitivity of XMCD. In Figure~\ref{fig4}, we present the XMCD spectra near the Ta N$_3$ absorption edge for a representative LaAlO$_3$/KTaO$_3$(111) sample at a temperature $T = $ 15 K under an in-plane magnetic field. Remarkably, the reproducible dichroism features appeared only near the absorption peaks (see Figure~\ref{fig4}b) are discernable that resemble XMCD signals were observed at LaAlO$_3$/SrTiO$_3$ heterointerfaces~\cite{JSLeeNM}, strongly indicating that these XMCD signals are indeed originating from the intrinsic spin polarization of the Ta $5d$ electrons. It is worth emphasizing that the magnetic behavior observed via Ta XMCD aligns with the SQUID measurements presented in Figure~\ref{fig2}.

To further control our experiments, we also re-prepare and post-anneal the LaAlO$_3$/KTaO$_3$(111) under various oxygen partial pressures (P$_{\mathrm{O2}}$) to effectively modulate the oxygen vacancies, thereby strengthening the theoretical scenario of emergent ferromagnetism induced by these vacancies, as illustrated in Figure~\ref{fig5}. As expected from our intuitions, the increase in P$_{\mathrm{O2}}$ gradually fills oxygen vacancies at the KTaO$_3$ heterointerfaces. This results in the enhancement of R$_{\mathrm{s}}$ (see Figure~\ref{fig5}a) in the two-dimensional electron gases, accompanied by a substantial reduction in both charge carrier density (see Figure~\ref{fig5}b) and ferromagnetization (see Figure~\ref{fig5}c,d). Notably, the magnetic response for the LaAlO$_3$/KTaO$_3$ post-annealed at a P$_{\mathrm{O2}}$ of 1 mbar is significantly diminished, exhibiting an almost vanishing magnetic hysteresis loop (see Figure~\ref{fig5}d). This underscores the pivotal role of oxygen vacancies in the emergence of interfacial ferromagnetism. Consequently, these independent and complementary electrical transport and magnetization measurements provide a robust and compelling evidence for that the ferromagnetic Ta at the superconducting KTaO$_3$ heterointerfaces is distinctly linked to the partially filled 5$d$ electrons brought about by the oxygen vacancies.

Physically, since the conducting bands at the Fermi energy at the KTaO$_3$ heterointerfaces are predominantly Ta 5$d$ orbitals, and the distinct ferromagnetism and superconductivity derive from the same Ta 5$d$ electrons both theoretically and experimentally (see Figures~\ref{fig3} and~\ref{fig4}), it is suggested that the superconductivity at the KTaO$_3$ heterointerfaces could spatially coexist with the ferromagnetism at low temperatures. In addition, a notable observation in Figure~\ref{fig2} is the two-dimensional superconductivity developed inside a ferromagnetic phase at the KTaO$_3$ heterointerfaces, implying that the ferromagnetism likely plays a crucial role in mediating the superconducting Cooper pairs~\cite{Ghosh}. Significantly, the KTaO$_3$ heterointerfaces reveal mixed-parity superconductivity with an admixture of $s$-wave and $p$-wave pairings~\cite{ref26}, compatible with the coexisting ferromagnetism. Moreover, it is intriguing that the KTaO$_3$ heterointerfaces exhibit the highest $T_c$ of superconductivity among ferromagnetic EuO overlayer~\cite{ref23,Hua2022}, suggesting that ferromagnetism may reinforce superconductivity. As a consequence, these complementary findings support the coexistence of ferromagnetism and superconductivity at the KTaO$_3$ heterointerfaces (see also Figure S14). The superconductivity observed at the KTaO$_3$ heterointerfaces is thus likely to be of nontrivial nature with broken time-reversal symmetry~\cite{Stewart}, making it promising for potential applications in topological quantum computing~\cite{Nayak,Alicea,Beenakker,Elliott}.

\notag\

\noindent\textbf{CONCLUSION}

\noindent In summary, we have experimentally developed two-dimensional superconducting heterointerfaces of KTaO$_3$ and uncovered remarkable in-plane ferromagnetic hysteresis loops that persists above room temperature. Additionally, we conduct the first-principles calculations and XMCD measurements to elucidate further the intrinsic ferromagnetism emerging at the heterointerfaces of KTaO$_3$. These intriguing findings highlight KTaO$_3$ heterointerfaces as time-reversal symmetry breaking superconductors. Our results therefore pave the way to further experimental and theoretical studies aimed at unraveling the intricate interplay among strong spin-orbit coupling, ferromagnetism, and superconductivity at the newly discovered KTaO$_3$ heterointerfaces.

\notag\

\noindent\textbf{Supporting Information}

\noindent The Supporting Information is available free of charge at online.

Methods; Large-area AFM images of thin films and KTaO$_3$ substrates; XRD images of thin films and KTaO$_3$ substrates; Anisotropic superconductivity of LaAlO$_3$/KTaO$_3$(111); Extrapolated BKT superconducting transition temperature; Temperature-dependent magnetization of KTaO$_3$ heterointerfaces; In-plane magnetization of KTaO$_3$ substrates; Out-of-plane magnetization of KTaO$_3$ substrates; XPS of KTaO$_3$ heterointerface and KTaO$_3$ substrate; SEM of KTaO$_3$ heterointerfaces; In-plane magnetization of LaAlO$_3$/KTaO$_3$(111); In-plane magnetization of YAlO$_3$/KTaO$_3$(111); Out-of-plane magnetization of LaAlO$_3$/KTaO$_3$(111); Out-of-plane magnetization of YAlO$_3$/KTaO$_3$(111); Field-dependent electrical characteristics of magnetization; High temperature-dependent in-plane magnetization for LaAlO$_3$/KTaO$_3$(111); Electronic and magnetic properties of KTaO$_3$ heterointerfaces; Calculated PDOS of (LaAlO$_3$)$_6$/(KTaO$_3$)$_6$(111) with an oxygen vacancy; Effective $U_{\mathrm{eff}}$ dependence of magnetic moments.

\notag\

\noindent\textbf{AUTHOR INFORMATION}

\noindent\textbf{$^*$Corresponding Author}

\textbf{Shan Qiao}, Email: qiaoshan@mail.sim.ac.cn

\textbf{Gang Mu}, Email: mugang@mail.sim.ac.cn

\textbf{Yan Chen}, Email: yanchen99@fudan.edu.cn

\textbf{Wei Li}, Email: w$\_$li@fudan.edu.cn

\notag\

\noindent\textbf{Author Contributions}

\noindent
Z.N., J.Q., and Y.L. contributed equally to this work. W.L. and Y.C. conceived the project and designed the experiments. Z.N. and G.Z. grew the samples. Z.N., Y.L., F.C., G.Z., and G.M. performed the electrical transport measurements. Z.N. and Y.L. performed the magnetization measurements. M.Z., L.D., and S.Q. performed the XMCD measurement. X.Y. and Q.G. performed the XPS measurement. X.Y., Q.G., and H.J. performed the SEM measurement. W.P. and Y.C. assisted the magnetization, electrical transport and SEM experiments. J.Q. performed the first-principles calculations. W.L. wrote the paper with input from Z.N. and J.Q.. All authors discussed the results and gave approval to the final version of the manuscript.

\notag\

\noindent\textbf{Notes}

\noindent
The authors declare no competing financial interest.

\notag\

\noindent\textbf{Acknowledgements}

\noindent
This work is supported by the National Natural Science Foundation of China (Grant Nos. 11927807 and U2032207) and Shanghai Science and Technology Committee (Grant Nos. 23ZR1404600 and 20DZ1100604). The authors also thank beamline BL07U of the Shanghai Synchrotron Radiation Facility (SSRF).

\end{document}